\documentclass[a4paper,12pt,DIV11]{scrartcl}
\usepackage[
	showeqnr,
	theoremdefs,
	 final
]{latexdev}

\usepackage[numbers]{natbib} 

\usepackage{subfig}
\usepackage[percent]{overpic}

\usepackage[affil-sl]{authblk} 

\title{Symmetry reduction for central force problems}
\author[1]{Robert McLachlan\thanks{\href{mailto:r.mclachlan@massey.ac.nz}{r.mclachlan@massey.ac.nz}}}
\author[2]{Klas Modin\thanks{\href{mailto:klas.modin@chalmers.se}{klas.modin@chalmers.se}}}
\author[3]{Olivier Verdier\thanks{\href{mailto:olivier.verdier@math.uib.no}{olivier.verdier@math.uib.no}}}

\affil[1]{
	Institute of Fundamental Sciences, Massey University, New Zealand
}
\affil[2]{
	Department of Mathematical Sciences, Chalmers University of Technology and the University of Gothenburg, Sweden
}
\affil[3]{
	Department of Mathematics, Bergen University College, Norway
}
\date{\today}

\def\R{\mathbb{R}}

\begin{document}
\maketitle
\begin{center}
\em Dedicated to the memory of Jerry Marsden, 1942--2010
\end{center}

\section{Introduction}

Euler's equations for the motion of a free rigid body exemplify the power of the geometric viewpoint in classical mechanics and provide a first look into the merits and subtleties of symmetry. 
Indeed, Euler's rigid body equations are a perennially popular topic in mechanics, as they can be introduced and studied with little prior knowledge, 
yet present an array of advanced phenomena and techniques. 
They illustrate nearly all aspects of geometric mechanics and are repeatedly revisited each time new ideas are encountered.
But, the almost exclusive focus on these equations does seem a little disproportionate. 
Why are the rigid body equations not one example amongst many, or, at least, the first in a sequence of increasing complexity and new phenomena? 

Part of the reason is that the general theory of Hamiltonian systems with symmetry,
exemplified by the rigid body equations, 
is thorny and still not fully understood. 
The quotient spaces that arise may have complicated topology and may be hard to work with; they may have different dimensions and fit together in tangled ways; they may fail to be manifolds. 
These complexities already occur in apparently simple situations, such as symmetries associated with the natural action of a matrix group on a vector space. 
Even the simplest general result, the \emph{Marsden--Weinstein--Meyer symplectic reduction theorem}, is a topic for an advanced course in geometric mechanics, while attempts to cover more complicated, singular group actions, such as Ortega and Ratiu \cite{or-ra}, are formidable and yet still incomplete.

In his lectures on physics~\cite[I, Ch. 22--1]{feynman} Feynman rhetorically comments on why he has a chapter on algebra:

{\medskip\narrower\narrower\em
Another reason for looking more carefully at algebra now, even though most of us studied algebra in high school, is that it was the first time we studied it; all the equations were unfamiliar, and it was hard work, just as physics is now. 
Every so often it is a great pleasure to look back to see what territory has been covered, and what the great map or plan of the whole thing is.
Perhaps some day somebody in the Mathematics Department will present a lecture on mechanics in such a way as to show what it was we were trying to learn in the physics course!

\medskip

}

\setlength{\leftskip}{0cm}

In this paper we take a closer look at Hamiltonian symmetry reduction applied to equations that are intimately familiar to every physicist.
First the rigid body equations, and then another example for which a concrete, elementary symplectic reduction can be carried out: the motion of a particle under a central force. 
This example duplicates and reinforces many of the features of the rigid body equations, and, like them, describes a physical system of fundamental importance. 
At the same time, the central force example provides a glimpse of a new structure of widespread importance in mathematical physics, namely the \emph{dual pair}. 
Somewhat surprisingly, given that central force problems are invariant under rotations, the Lie algebra that arises is not $\mathfrak{so}(3)$, the antisymmetric $3\times 3$ matrices%
\footnote{All groups and algebras in this paper are real; thus $\mathfrak{so}(3)=\mathfrak{so}(3,\R)$,
$\mathfrak{sp}(2)=\mathfrak{sp}(2,\R)$, etc.}%
, but $\mathfrak{sp}(2)$, the  $2\times 2$ matrices of zero trace. 
As these two are the most important 3-dimensional Lie algebras, the study of the ``reduced central force equations'' provides an important balancing perspective to the rigid body equations.

\section{The rigid body equations}

\begin{figure}
\begin{center}
\begin{overpic}[width=0.7\textwidth]{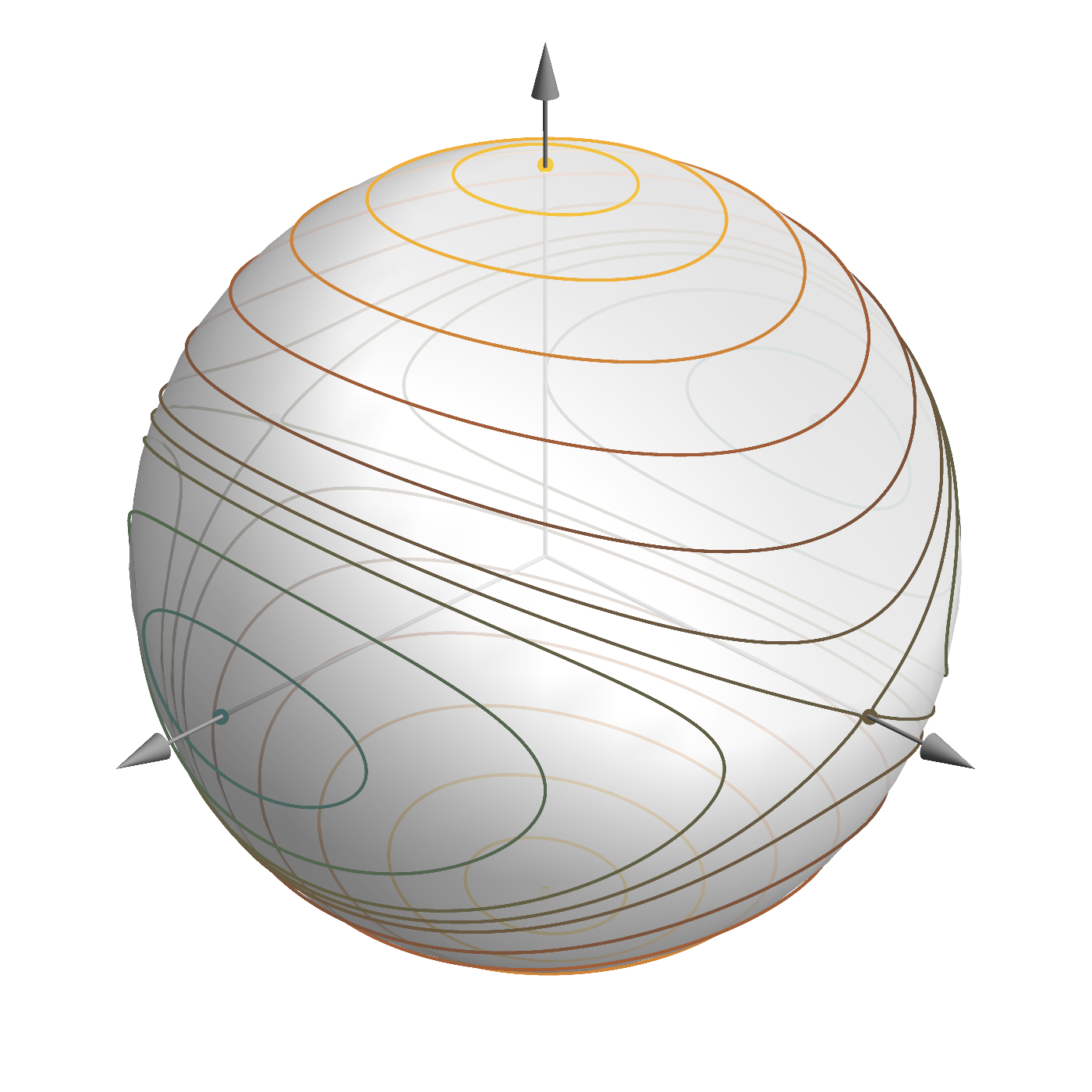}
\put (5,25) {$m_1$}
\put (92,25) {$m_2$}
\put (42,96) {$m_3$}
\end{overpic}
\caption{\label{fig:phaseportrait}
Phase portrait of the free rigid body. 
The sphere represents one level set of the angular momentum~$C(m)$.
The curves on the sphere represent several level sets of the Hamiltonian~$H(m)$ restricted to the angular momentum sphere.
Because both the angular momentum and the Hamiltonian are conserved, these level sets constitute the phase portrait.
Notice that there are 6 relative equilibria: $(\pm m_1,0,0)$, $(0,\pm m_2,0)$, and $(0,0,\pm m_3)$.
Each pair correspond to rotation, in positive or negative direction, about a principal axis.
From the diagram one directly makes out that rotations about the $m_1$ and $m_3$ axes are stable, whereas rotations about the $m_2$ axis are unstable.
The hammer throw experiment, illustrated in Fig.~\ref{fig:hammer}, is an easy way to demonstrate this phenomenon.
}
\end{center}
\end{figure}

\begin{figure}
\begin{center}
\includegraphics[width=10cm]{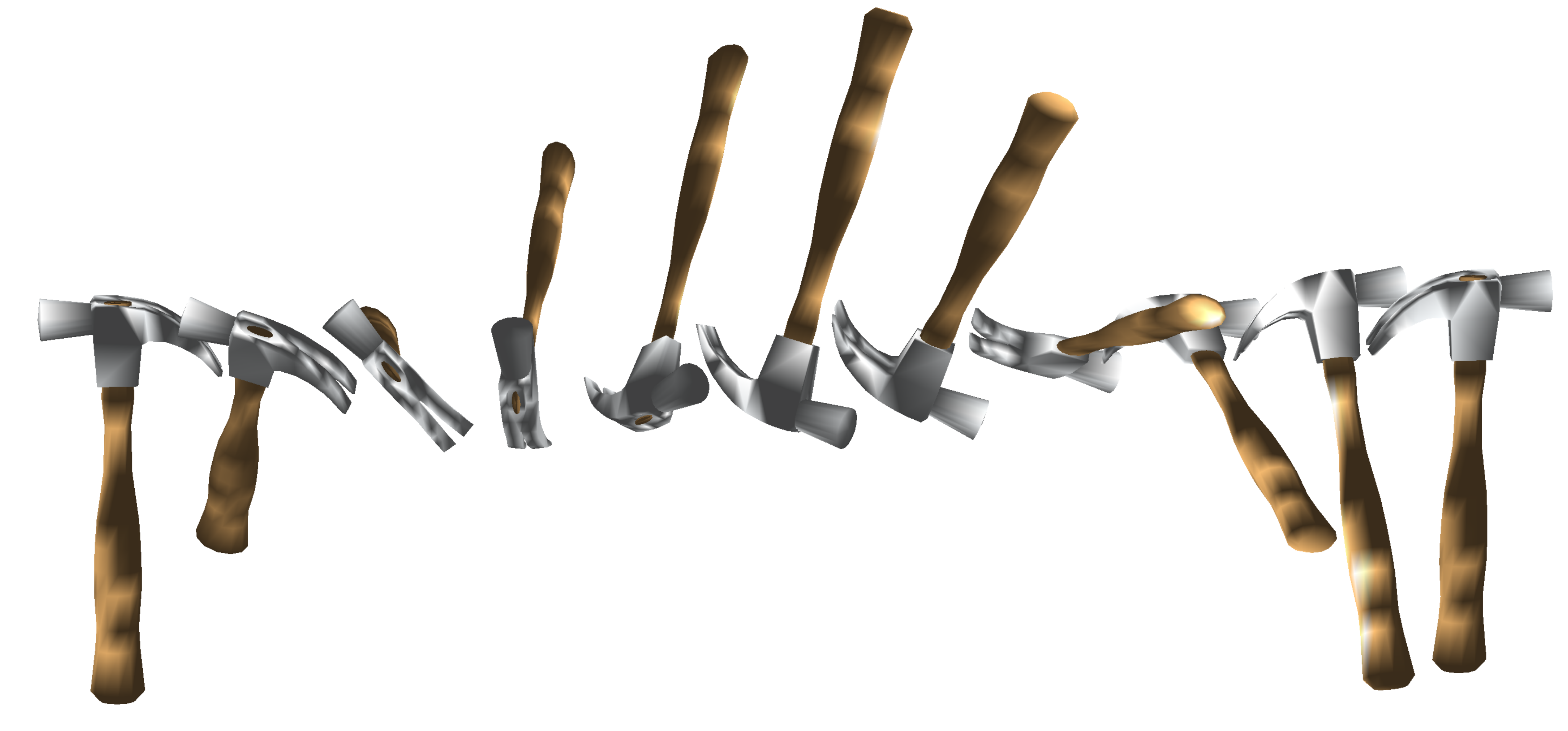}
\includegraphics[width=10cm]{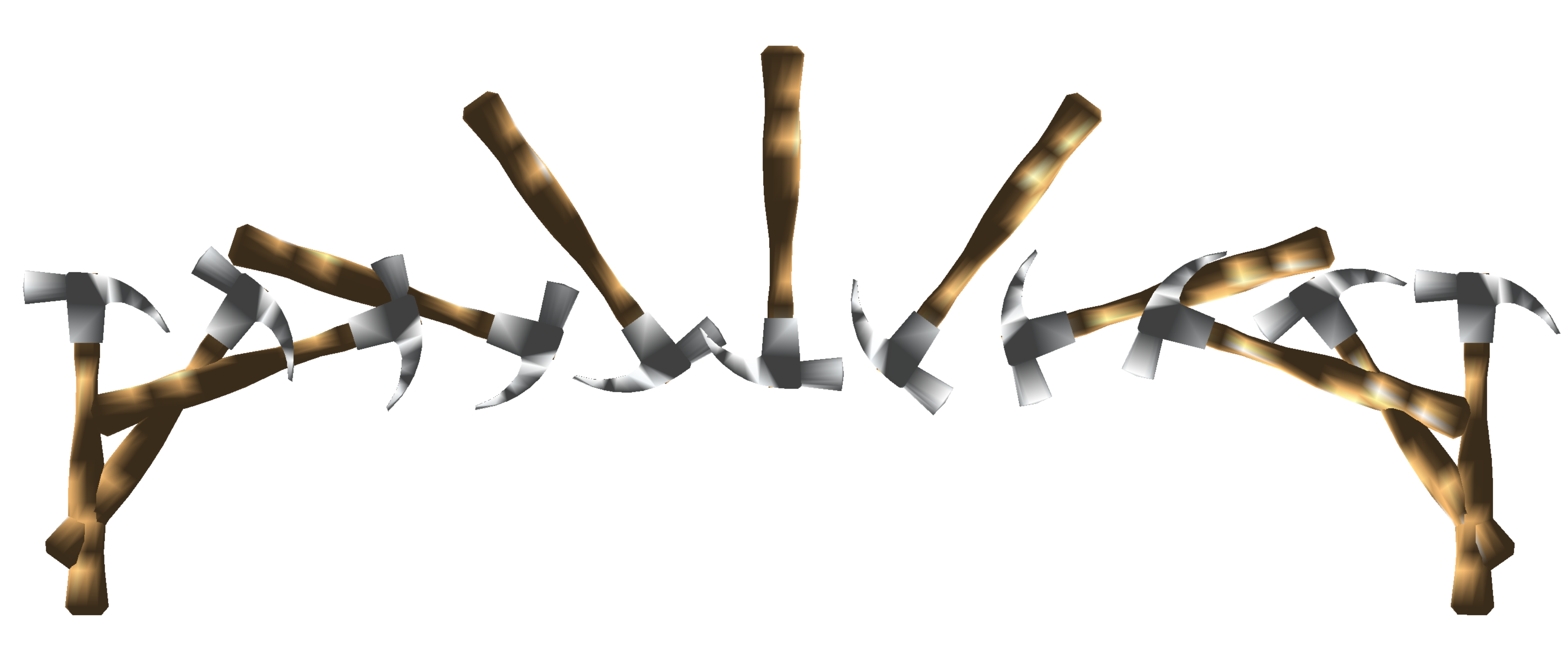}
\includegraphics[width=10cm]{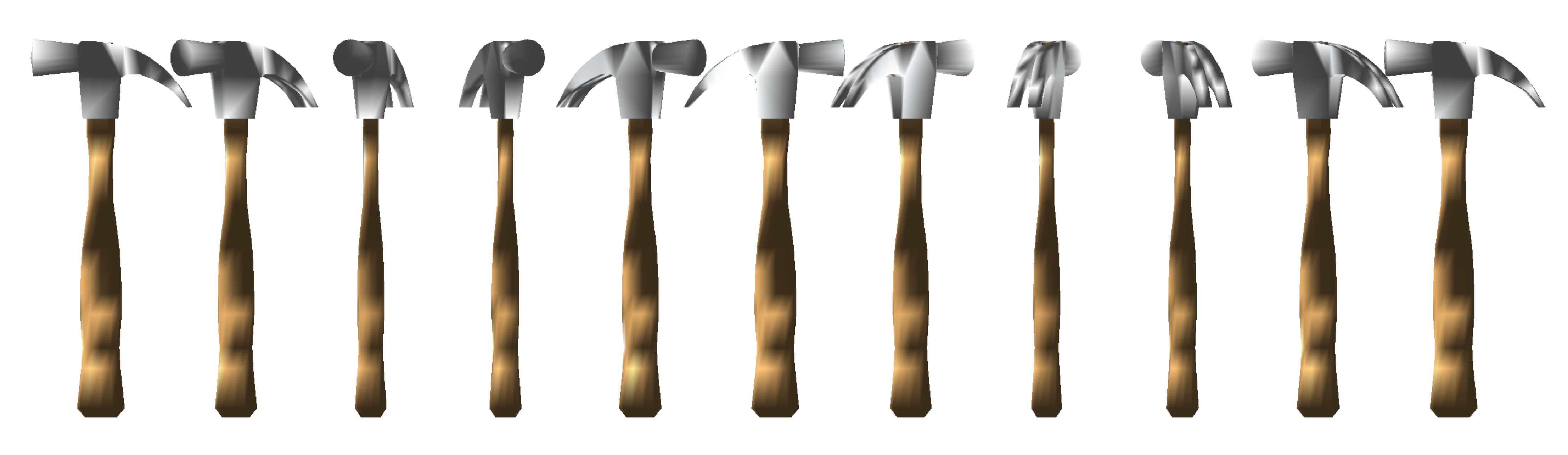}
\caption{\label{fig:hammer} Simulation of the hammer throw experiment. The rotation of a hammer is shown when it is launched with an initial rotation about each of its three principal axes. The top rotation is unstable, resulting in the hammer undergoing a flip and landing in the opposite orientation, while the other two are stable.
}
\end{center}
\end{figure}

The Euler equations for the motion of a triaxial free rigid body are
\begin{equation}
\begin{aligned}
\label{eq:euler}
\dot m_1 &= \left(\frac{1}{I_3}-\frac{1}{I_2}\right) m_2 m_3 \\
\dot m_2 &= \left(\frac{1}{I_1}-\frac{1}{I_3}\right) m_3 m_1 \\
\dot m_3 &= \left(\frac{1}{I_2}-\frac{1}{I_1}\right) m_1 m_2, \\
\end{aligned}
\end{equation}
where $I_1\geq I_2\geq I_3 > 0$ are the principal moments of inertia of the body, and $m_i$ is the angular momentum about the $i$th principal axis of the body. 
The total angular momentum
\begin{equation}
C(m) \coloneqq m_1^2+m_2^2+m_3^2
\end{equation}
and the kinetic energy
\begin{equation}
\label{eq:H}
H(m) \coloneqq \frac{1}{2}\left( \frac{m_1^2}{I_1} +  \frac{m_2^2}{I_2} +  \frac{m_3^2}{I_3}\right)
\end{equation}
are conserved quantities.
Consequently, the phase portrait of \eqref{eq:euler} is given by 
the intersection of the spheres $C=$ const. and the ellipsoids $H=$ const., as shown in Fig.~\ref{fig:phaseportrait}.
This phase portrait shows that when the moments of inertia are distinct, i.e. when $I_1>I_2>I_3$, rotation about the $m_1$ and $m_3$ axes (the fixedÁ points $(\pm m_1,0,0)$ and $(0,0,\pm m_3)$, respectively) are stable, while rotation about the intermediate $m_2$ axis (the intermediate moment of inertia axis) is unstable. 
Indeed, rotations that start near $(0,m_2,0)$ will follow the heteroclinic orbit to a neighbourhood of $(0,-m_2,0)$. 
This can be illustrated in practice using the famous hammer throw experiment, shown in Fig.~\ref{fig:hammer}. 
In fact the experiment illustrates even more; in the course of traversing the heteroclinic orbit, the attitude of the hammer undergoes a rotation by $\pi$---an example of a geometric (or Hannay--Berry) phase.\footnote{%
Many people know this trick, but we suspect that not many could explain why the rotation angle is $\pi$. For a complete explanation of why a rotation by $\pi$ is observed for `hammer-like' objects, see
Cushman and Bates \cite[III.8]{cu-ba}.}

\begin{figure}
\begin{center}
\includegraphics[width=6cm]{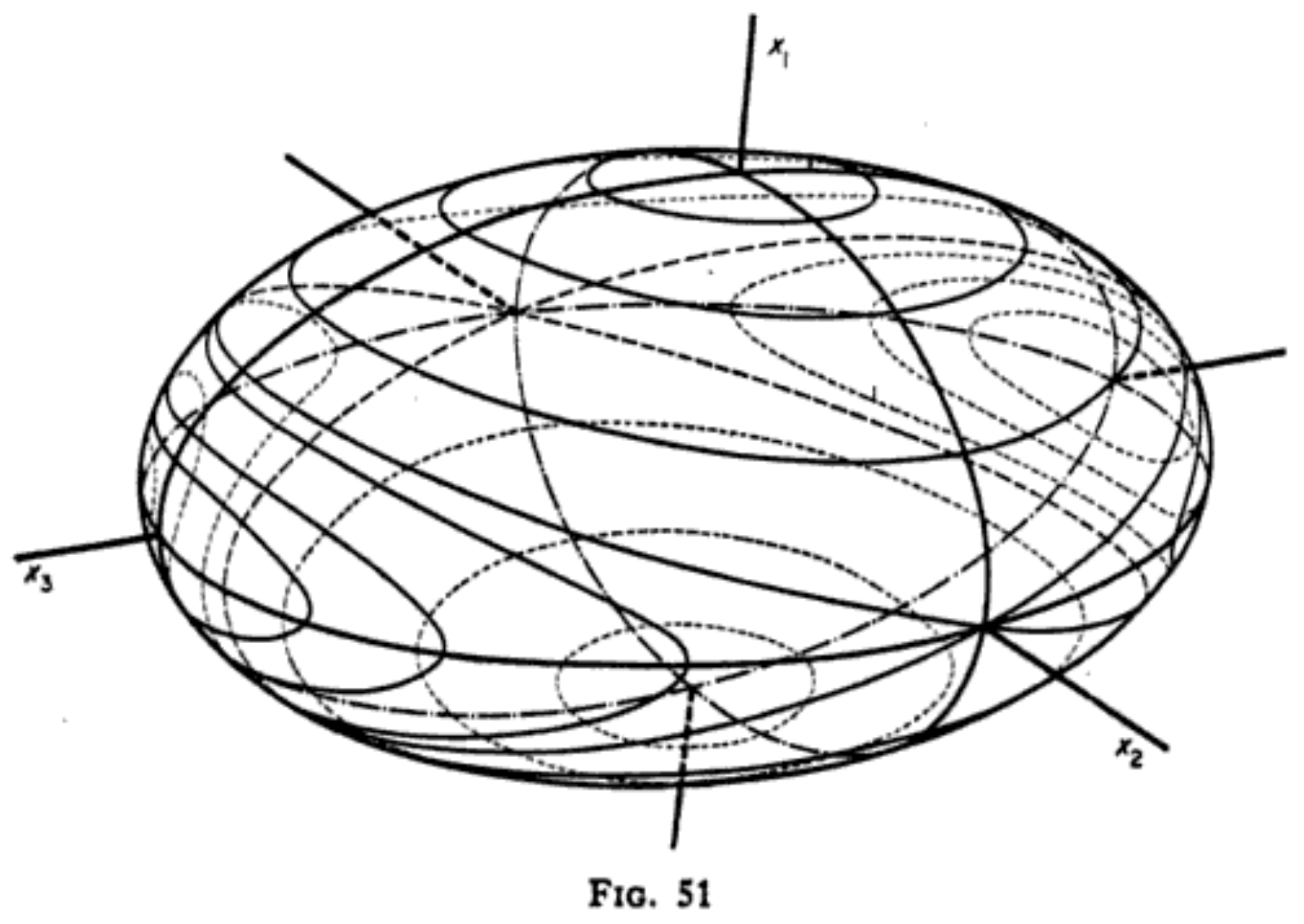}
\caption{\label{fig:ll}
The phase portrait of the Euler equations of the free rigid body as drawn
in Landau and Lifshitz's {\em Mechanics} \cite{la-li}.
}
\end{center}
\end{figure}

\setlength{\leftskip}{1.5cm}
{\medskip \small 

Although most undergraduate physics texts derive the Euler equations, few show the phase portrait 
as in Fig.~\ref{fig:phaseportrait}. Usually, an analytic approach is used to study small oscillations
about the stable rotation axes or to consider only symmetric rigid bodies, those  with $I_1=I_2$. One
classic physics text that does include the phase portrait is Landau and Liftshitz, {\em Dynamics} \cite{la-li},
first published in Russian in 1940 and in English in 1960. Their version
is shown in Fig.~\ref{fig:ll}. Another (although hardly a standard undergraduate text) is Arnold's 
{\em Mathematical Methods of Classical Mechanics} \cite{arnold2}---see Fig. 121. (Arnold
draws a constant-energy ellipsoid, instead of a constant-angular-momentum sphere.)
A mathematics text that includes the phase portrait is 
Bender and Orszag, {\em Advanced Mathematical Methods for Scientists and Engineers}, Fig. 4.31 \cite{be-or}.

\medskip

}

\setlength{\leftskip}{0cm}

A more detailed treatment would view Eq.~(\ref{eq:euler}) as the result of a {\em symmetry reduction}.   
Indeed, symmetry has a claim to be the single most important unifying and organizing principle in physics.
The {\em attitude} of a free rigid body is specified by its rotation from an initial reference position. 
Thus, its configuration space is $\mathrm{SO}(3)$, the proper $3\times3$ orthogonal matrices. 
Its phase space is the cotangent bundle $T^*\mathrm{SO}(3)$, the space of possible attitudes
and angular momenta of the body. It's convenient to work with the closely
related space $\mathrm{SO}(3)\times\mathfrak{so}(3)^*$, whose points can be written as $(Q,m)$ where $Q\in\mathrm{SO}(3)$ is the orientation of the body and $m\in\mathfrak{so}(3)^*\approx \R^3$  is its angular momentum in the ``body frame'', that is, in coordinates that are fixed in the body and rotate along with it.%
\footnote{A more classical formulation, originally by Euler, is to use 3 Euler angles for the attitude together with 3 conjugate momenta---6 variables altogether. However, the Euler angles do not cover all possible attitudes of the body and are singular near some attitudes.}
The full, unreduced equations of motion are then 
\begin{align}
\label{eq:full1}
\dot Q &= Q\left( \begin{matrix}
0 & -\omega_3 & \omega_2 \\
\omega_3 & 0 & -\omega_1 \\
-\omega_2 & \omega_1 & 0 \\
\end{matrix}
\right),
\\
\label{eq:full2}
\dot m &= m\times \omega,
\end{align}
where $I_j \omega_j = m_j$, $j=1,2,3$, and
$\omega\in\mathfrak{so}(3)\approx\R^3$ is the angular velocity in the body frame.

\def\M{\mathcal{M}}

The symmetry group here is $\mathrm{SO}(3)$ and it acts on phase space by
$A\cdot (Q,m) = (AQ,m)$; the system (\ref{eq:full1},\ref{eq:full2}) is invariant under this group. 
As a consequence, the system has 3 conserved quantities associated with the symmetry (the spatial angular momentum $Qm$)
and a {\em reduced} subsystem given by Eq.~(\ref{eq:full2}) (or equivalently, (\ref{eq:euler})).

More generally, symmetry in classical mechanics enters via the study of \emph{Hamiltonian systems with symmetry}
(see \cite{ma-ra} for an introduction to this area, and \cite{ma-we2} for its history).
Such systems consists of (i) a phase space $\M$ equipped with a symplectic form, (ii) a symmetry group $G$ acting on $\M$, (iii) a momentum map $J$, and (iv) a $G$-invariant Hamiltonian $H$.
(For the rigid body, these data are $\M=T^*\mathrm{SO}(3)$; $G=\mathrm{SO}(3)$ with action $A\cdot(Q,m)=(AQ,m)$;
$J(m)=Qm$, the spatial angular momentum; and $H(m)$ as given in Eq. (\ref{eq:H}).)
These data $(\M ,G,J,H)$ interact in complicated ways. 
In many cases, they allow great insight into the behaviour of the system, such as the construction of some or all orbits and their stability and bifurcations. Many phenomena are independent of the specific Hamiltonian $H$, so that the `geometric' data $(\M ,G,J)$ defines a class of dynamical systems all of whose members share typical dynamical features.

Specifically, the two foliations of $\M $ into level sets of $J$ and into orbits of $G$ are both preserved by the flow of systems associated to $G$-invariant Hamiltonians. The first of these is just the statement that the momentum
$J$ is a conserved quantity---Noether's theorem. 
That is, each momentum level set $J(m)=$const. is invariant under the flow.
The second foliation is more subtle as the flow takes one group orbit into another. In addition, both foliations interact with the symplectic structure. Symplectic reduction allows one to reduce the dynamics to a new phase space of smallest possible dimension, and reconstruction lifts the reduced dynamics back to the original phase space. A key result, for example, is the celebrated Marsden--Weinstein--Meyer theorem \cite{ma-we,meyer}.

The rigid body equations (\ref{eq:euler}) also offer a popular way to motivate the introduction of noncanonical Hamiltonian systems. Canonical systems are even-dimensional and hence the 3-dimensional rigid body equations cannot
possibly be canonical. 
Instead, they form a {\em Poisson system}.
We recall that a Poisson bracket $\{,\}$ on $\M $ is an operation that satisfies the axioms
\begin{equation*}
\begin{aligned}
\{,\}& \colon C^\infty(\M )\times C^\infty(\M ) \to C^\infty(\M )\\
\{F,G\} &= -\{G,F\}\\
\{F,a G + b H\} &= a\{F,G\} + b\{F,H\}\\
\{F,GH\} &= \{F,G\}H + \{F,H\}G 
\end{aligned}
\end{equation*}
for all $F,G,H\in C^\infty(\M )$ and for all $a, b\in \mathbb{R}$. 
The Poisson system  with Hamiltonian $H$ is then
\begin{equation*}
\dot m = \{m,H\},\quad m\in \M .
\end{equation*}
In local coordinates $m$ on $\M $, Poisson systems can also be written in terms of the Poisson tensor $K$ as
\begin{equation}
\label{eq:poisson}
\dot m = K(m)\nabla H(m)
\end{equation}
where
\begin{equation*}
\{F,G\} = (\nabla F)^T K(m) (\nabla G).
\end{equation*}

In the case of the rigid body equations, 
\begin{equation}
\label{eq:M}
K(m) = \left( \begin{matrix}
0 & -m_3 & m_2 \\
m_3 & 0 & -m_1 \\
-m_2 & m_1 & 0 \\
\end{matrix}
\right).
\end{equation}
Poisson systems in which $\M $ is a vector
space and $K(m)$ is a linear function of $m$ play a special role. It is easy to see that
in this case $\M$ is the dual of a Lie algebra \cite{olver}; such an $\M$ is called 
{\em Lie--Poisson}. For the rigid body equations, $\M =\mathfrak{so}(3)^*$. The conserved
quantitity $C$ is now seen to be a {\em Casimir}, a function whose Poisson bracket with
all smooth functions is zero, and which is therefore conserved for all Hamiltonians. 
The Poisson manifold $\M $ is foliated into submanifolds called symplectic leaves,
each of which is a symplectic manifold; often they are the level sets of the Casimirs.
For the rigid body equations, the symplectic leaves are the spheres $C(m) = $ const.
They are also the motivating example of 
the whole class of {\em Euler--Arnold equations} that describe geodesics on groups with respect to an invariant metric. 
Such equations were advocated in 1966, when Arnold~\cite{arnold} discovered that the Euler rigid body and the Euler fluid equations share this basic structure, with the groups being $\mathrm{SO}(3)$ in the first case and the volume-preserving diffeomorphisms in the second case.

Thus, the rigid body equations provide an excellent example of
(i) systems with nonlinear phase spaces;
(ii) visualization of phase portraits;
(iii) noncanonical Hamiltonian systems;
(iv) Poisson systems;
(v) Lie--Poisson systems;
(vi)  Marsden--Weinstein--Meyer symplectic reduction;
(vii) reconstruction and geometric phases;
(viii) Casimirs;
(ix) Euler--Arnold systems;
(x) dynamical phenomena discovered using geometric methods, without solving any differential equations.
Moreover, they apply to readily accessible everyday experiences, notably the hammer throw experiment and the rotation of the earth.\footnote{A topic of current research interest; see, e.g.,\cite{schreiber}.}


\section{The reduced central force equations}
Consider a particle moving in 3 dimensions subject to a conservative central force. It has position $q\in\mathbb{R}^3$,
momentum $p\in\mathbb{R}^3$, and total (kinetic plus potential) energy $H(q,p) = \frac{1}{2}\|p\|^2 + V(\|q\|^2)$. The equations of motion are Hamilton's equations,
\begin{equation}
\label{eq:eom}
\begin{aligned}
\dot q &= \frac{\partial H}{\partial p} = p \\
\dot p &= -\frac{\partial H}{\partial q}=-2V'(\|q\|^2)q 
\end{aligned}
\end{equation}
The phase space is $\M = T^*\mathbb{R}^3\cong \mathbb{R}^6$ and the Hamiltonian is orthogonally invariant. 
That is, the symmetry group is $G=\mathrm{O}(3)$, the $3\times 3$ orthogonal matrices, which
acts on $\M$ by $A\cdot(q,p) = (Aq,Ap)$. The symmetry applies the same rotation $A$ to both the position
$q$ and the momentum $p$. The Hamiltonian $H(q,p)$ is invariant under the symmetry, i.e.,
$H(Aq,Ap) = H(q,p)$ for all orthogonal matrices $A$. The equations of motion, Eq.~(\ref{eq:eom}), 
are orthogonally invariant as well.

We want to factor out this symmetry group to obtain
reduced equations. As for the rigid body, this will be done in two steps:
\begin{itemize}
\item[(i)] by passing to the space of orbits of $G$, yielding a Lie--Poisson system;
\item[(ii)] by finding the Casimir of this system and hence passing to a symplectic leaf.
\end{itemize}
In fact, as this procedure can be followed for {\em any} $\mathrm{O}(3)$-invariant Hamiltonian, we will
carry it out for the general case.



If the Hamiltonian is $\mathrm{O}(3)$-invariant, that means it is constant on each orbit of the group.
The group action, a common rotation or reflection of $q$ and $p$, preserves $\|q\|^2$ and $\|p\|^2$, and it also preserves $q\cdot p$, essentially the angle between $q$ and $p$. It is easy to see that given any two pairs $(q,p)$ with the same
values of $\|q\|^2$, $q\cdot p$, and $\|p\|^2$, there must be a symmetry operation---a common
orthogonal transformation of $q$ and $p$---that sends one
pair into the other. That is, we can use the values of
\begin{equation}
\label{eq:w}
\begin{aligned}
w_1 & := \|q\|^2,\\
w_2 &:= q\cdot p,\\
w_3 &:= \|p\|^2.
\end{aligned}
\end{equation}
to label the different group orbits. We have thus found that any $\mathrm{O}(3)$-invariant Hamiltonian, $\bar H(q,p)$ say,
can be written in the form
\begin{equation}
\label{eq:H2}
\bar H(q,p) = H(\|q\|^2,q\cdot p,\|p\|^2)
\end{equation}
for some function $H$. 

\begin{figure}
\centering
 \begin{center}
 \includegraphics[width=0.7\textwidth]{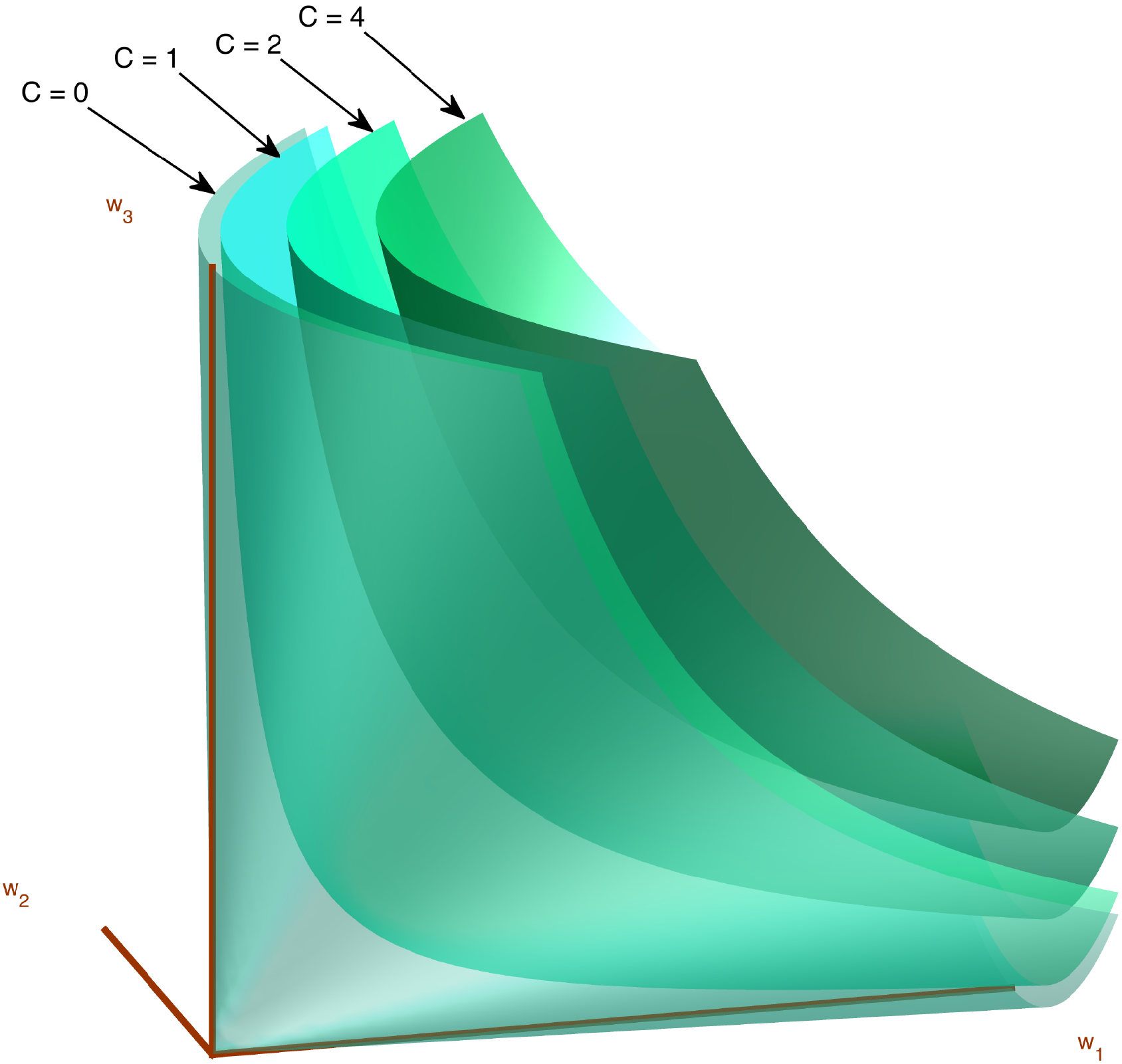}
\caption{\label{fig:coadjoint}
The reduced phase space of central force problems. Here $w_1=\|q\|^2$, $w_2=q\cdot p$, and $w_3=\|p\|^2$
are the reduced coordinates, and $C=w_1 w_3 - w_2^2$ is the Casimir corresponding to
angular momentum. The motion is restricted to level sets of the Casimir $C$. These
are the symplectic leaves or symplectic reduced spaces for central force problems. There
are three types of leaves: for $C>0$ the leaves are one sheet of a 2-sheeted hyperboloid
(topologically a plane), and for $C=0$ (i.e., for zero angular momentum) they are the single point at the origin and
the cone meeting the origin (topologically a cylinder).
All orbit types fit smoothly together.
}
\end{center}
\end{figure}

\setlength{\leftskip}{1.5cm}
{\medskip \small 

The appearance of $w_1$, $w_2$, and $w_3$ is no surprise, as the orthogonal group $\mathrm{O}(3)$ can be {\em defined} to be the linear maps that preserve all dot products between vectors. However, although we
reached (\ref{eq:H2}) easily, there is much more going on here, which we will amplify in this aside.

The quadratics $\|q\|^2$, $q\cdot p$, and $\|p\|^2$ are {\em invariants} of the symmetry group and any function of invariants, like (\ref{eq:H2}), is automatically invariant. But much more is true. It is known for this symmetry group that {\em any} invariant polynomial in $q$, $p$ must be a polynomial in the $w_1$, $w_2$, and $w_3$. This is an example of what is called a {\em First Fundamental Theorem} for a symmetry group; such a theorem is known in only a few cases. From there it is a small step to conjecture that if $\bar H(q,p)$ is a smooth invariant function, then $H$ will be a smooth function as well. This turns out to be true.  ({\em A priori}, one might have worried the $w_i$ being quadratic functions of $q$ and $p$ might mean $H$ could end up involving square roots. That can't happen.) This is an example of a very general and celebrated theorem of Schwartz \cite{schwarz} that states that for a compact group acting on a vector space, {\em any} smooth invariant function on the vector space is a smooth function of the invariants. In fact, it must be a smooth function of the polynomials that generate the invariant polynomials. In our case these generating polynomials are $\|q\|^2$, $q\cdot p$, and $\|p\|^2$.

While Hilbert's invariant theorem guarantees that for many (but not all) matrix
groups the set of invariant polynomials is finitely generated, it may not be a simple manner
to actually construct a generating set. 

Furthermore, we showed above why the invariants in this example serve to uniquely label the 
orbits of the symmetry group. In other examples, even when a generating set for the invariants can be found,
those invariants may not serve to uniquely label the the orbits. For example, the 1-dimensional scaling group acting
on the plane has no continuous invariants, because the orbits (open rays) have
a common limit point, the origin. So the situation considered here is
actually quite special.
\medskip

}

\setlength{\leftskip}{0cm}

For the Hamiltonian $H(\|q\|^2,q\cdot p,\|p\|^2)=H(w_1,w_2,w_3)$, 
Hamilton's equations are
\begin{equation}
\label{eq:central}
\begin{aligned}
\dot q &= \frac{\partial H}{\partial p}\hspace{3mm} = q \frac{\partial H}{\partial w_2} + 2p \frac{\partial H}{\partial w_3}, \\
\dot p &= -\frac{\partial H}{\partial q} = -2q \frac{\partial H}{\partial w_1} - p \frac{\partial H}{\partial w_2}.
\end{aligned}
\end{equation}

As $\|q\|^2\ge 0$, $\|p\|^2\ge 0$, and $\|q\times p\|^2 = \|q\|^2 \|p\|^2 - (q\cdot p)^2 \ge 0$, not all
values of $(w_1,w_2,w_3)$, which label the different group orbits, are realizable. The space
of group orbits is contained in the cone
\begin{equation*}
\left\{(w_1,w_2,w_3)\in\mathbb{R}^3\colon w_1\ge 0, w_3\ge 0, w_1 w_3 - w_2^2\ge 0\right\}
\end{equation*}
and it is easy to check that it does consist of the entire cone.
This situation, in which we have a complete description of the space of group orbits, with each orbit represented as a point in a vector space, is quite rare.

As we know that the flow of any $G$-invariant system, Hamiltonian or not, maps orbits to orbits, it is no surprise that the $w_i$ obey a closed subsystem, which can be computed using nothing more than the chain rule:
\begin{equation}
\label{eq:centralred}
\begin{aligned}
\dot w_1 &= 2 q\cdot \dot q & = \hphantom{-}2 w_1 \frac{\partial H}{\partial w_2} + 4 w_2 \frac{\partial H}{\partial w_3}, \\
\dot w_2 &= q\cdot \dot p + \dot q \cdot p  &= -2w_1 \frac{\partial H}{\partial w_1} + 2w_3 \frac{\partial H}{\partial w_3},\\
\dot w_3 &= 2 p \cdot \dot p & = -4 w_2 \frac{\partial H}{\partial w_1} - 2 w_3 \frac{\partial H}{\partial w_2}.
\end{aligned}
\end{equation}
Probably any student confronted with (\ref{eq:central}) and comfortable with the chain rule would hit on the idea
of computing (\ref{eq:centralred}). However, from the general theory of symplectic reduction we know 
that we have carried out the reduction $(T^*\R^3)/\mathrm{O}(3)$ and that the resulting system
(\ref{eq:centralred}) will be Poisson. 

Writing (\ref{eq:centralred}) explicitly as a Poisson system---that is, in the form of Eq. (\ref{eq:poisson})---yields
\begin{equation}
\label{eq:centralpoisson}
\begin{aligned}
\dot w &= \left(\begin{matrix}
0 & 2 w_1 & 4w_2 \\[1mm]
-2 w_1 & 0 & 2w_3 \\[1mm]
-4 w_2 & -2 w_3 & 0 \\
\end{matrix}
\right)
\left(
\begin{matrix}
\displaystyle
\frac{\partial H}{\partial w_1}\\[4mm]
\displaystyle
\frac{\partial H}{\partial w_2}\\[4mm]
\displaystyle
\frac{\partial H}{\partial w_3}
\end{matrix}
\right)\\
&=: \hspace{17mm} K(w) \hspace{15mm} \nabla H(w).
\end{aligned}
\end{equation}
Note the analogy with the rigid body equations in Lie--Poisson form (Eqs. (\ref{eq:H}), (\ref{eq:poisson}), (\ref{eq:M})).
Only the Poisson tensor is different, and the Hamiltonian is arbitrary. Thus, the reduced central force equations
(\ref{eq:centralpoisson}) form another and very easily accessible example of a Poisson system.

The Poisson tensor $K(w)$ in (\ref{eq:centralpoisson}) is linear in $w$ so it is natural to ask which Lie 
algebra it is associated with. That is, we seek 3 matrices $W_1$, $W_2$, $W_3$ such that
\begin{equation*}
[W_1,W_2] = 2 W_1,\quad [W_1,W_3] = 4 W_2,\quad [W_2,W_3] = 2 W_3.
\end{equation*}
A little trial and error yields the solution
\begin{equation*}
W_1 = \left( \begin{matrix} 0 & 2 \\ 0 & 0 \end{matrix}\right),\quad
W_2 = \left( \begin{matrix} -1 & 0 \\ 0 & 1 \end{matrix}\right),\quad
W_3 = \left( \begin{matrix} 0 & 0 \\ -2 & 0 \end{matrix}\right)
\end{equation*}
which are a basis for the Lie algebra $\mathfrak{sp}(2)$ of $2\times 2$ matrices
of zero trace. We will come to the question of why this {\em particular} Lie algebra appears later, in Section \ref{sec:dualpairs}.

To draw the phase portrait of the reduced central force equations (\ref{eq:centralpoisson}),
we first seek a Casimir $C$, that is, a function such that $K(w)\nabla C \equiv 0$. The solution is
\begin{equation*}
C(w) = w_1 w_3 - w_2^2.
\end{equation*}
The symplectic leaves are the level sets of $C$, which are hyperboloids for $C>0$ and a cone minus its point for $C=0$.
(There is also a singular leaf consisting of the single point $(w_1,w_2,w_3)=(0,0,0)$.)
These are shown in Figure.~\ref{fig:coadjoint}.
Note that, when written in terms of the original variables $(q,p)$, $C$ is the square of the total
angular momentum, $C(w) = \|q\times p\|^2$---in close analogy with the rigid body.
The orbits of (\ref{eq:centralpoisson}) are the intersection of the symplectic leaves with the
level sets of the energy $H(w)$---again in analogy with the rigid body.
This allows the phase portrait to be read off, for any $\mathrm{O}(3)$-invariant Hamiltonian, for
all values of the total angular momentum at once, and to easily visualize how all the orbits fit together.

\begin{figure}
\centering
\includegraphics[width=0.5\textwidth]{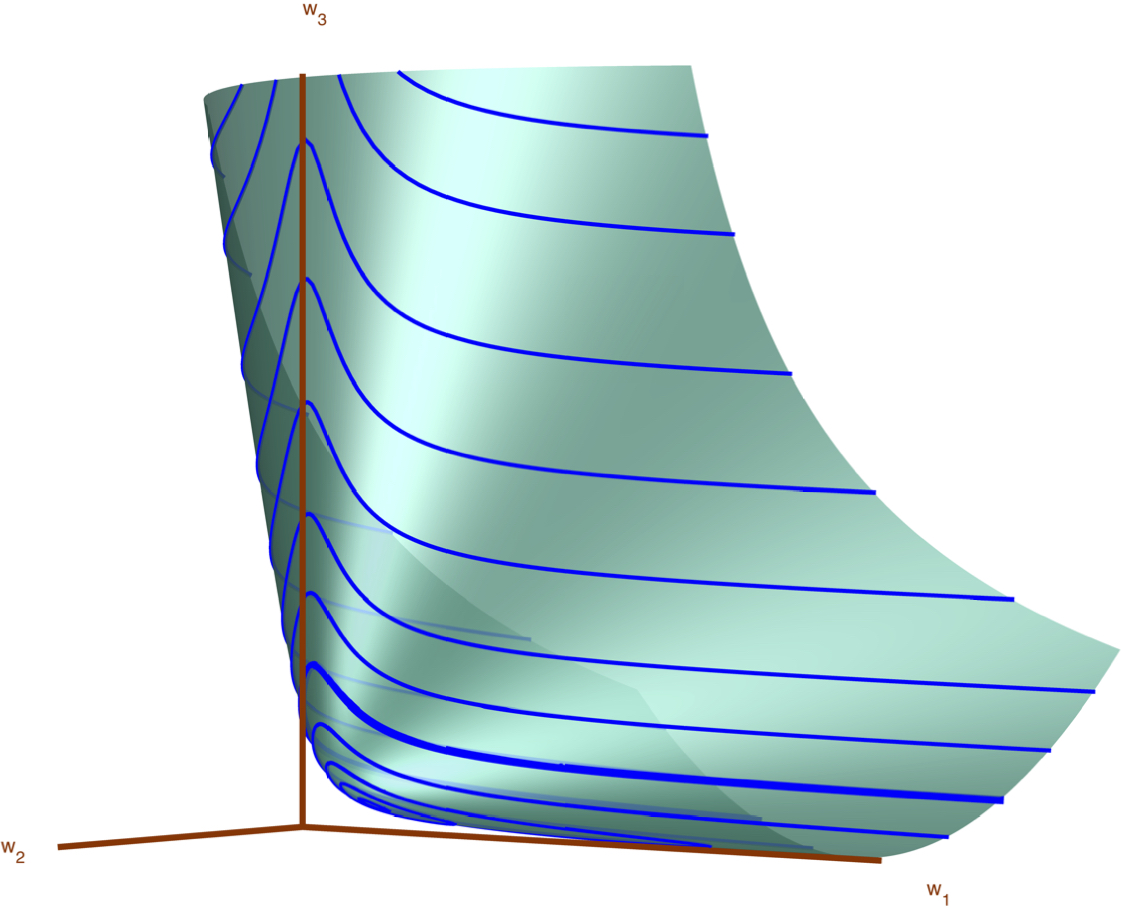}
\includegraphics[width=0.4\textwidth]{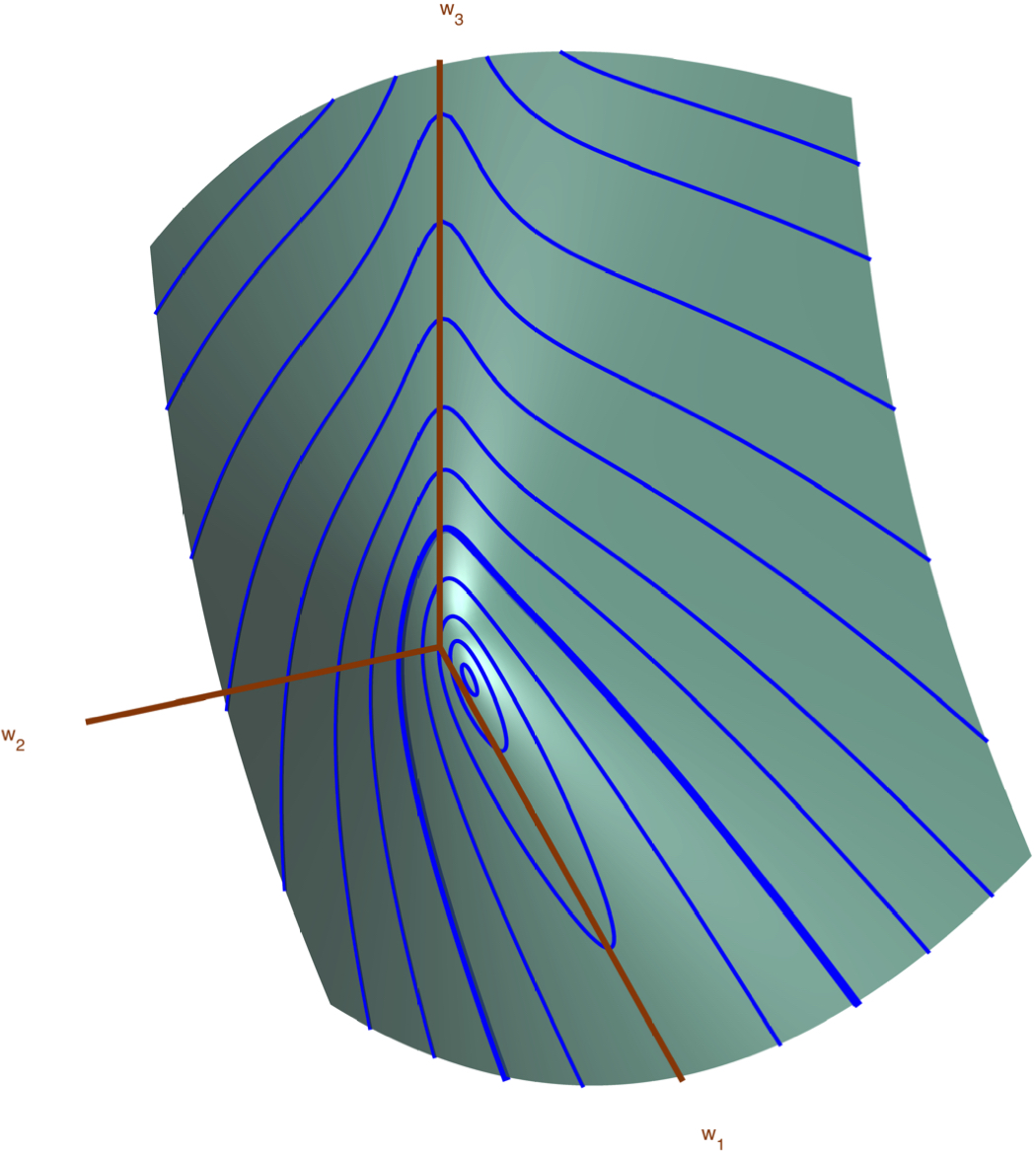}
\caption{\label{fig:coloumb}
Two views of the phase portrait for the reduced Kepler problem, given as level sets of the Hamiltonian $H$ restricted to the coadjoint orbit $C=0.6$.
The level set marked by a thicker curve corresponds to zero energy $H=0$. 
This is the \emph{escape energy}, so orbits below it are bounded and periodic, whereas curves above it are unbounded.
Notice the stable relative equilibrium at the bottom, corresponding to the circular solution of the Kepler problem where $\|q\|^2$ and $\|p\|^2$ are constant and $q\cdot p=0$. The region $[0,12]\times [-6,6]\times [0,12]$ is shown.
}
\end{figure}

\begin{figure}
\centering
\includegraphics[width=0.5\textwidth]{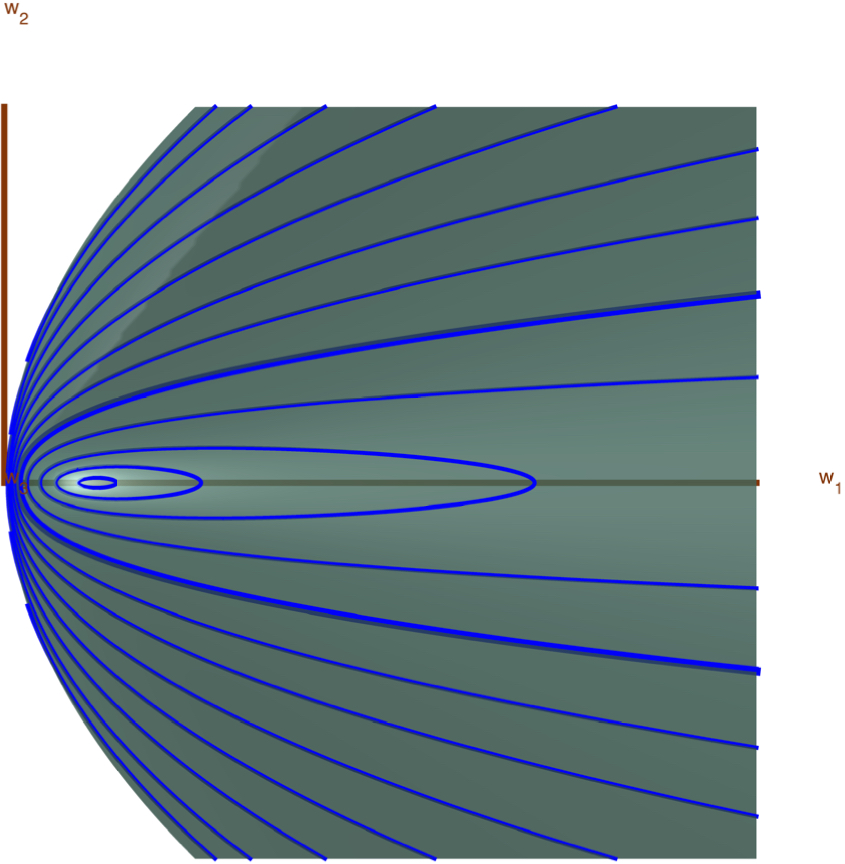}
\caption{\label{fig:coulomb2}
A more standard choice of reduced coordinates is $(w_1,w_2)$. The phase portrait of the reduced Kepler problem for $C=0.6$  is shown here in these coordinates. A large part of the phase portrait is 
squashed near the $w_2$-axis, and these coordinates cannot be used for zero angular momentum
$(C=0)$.
}
\end{figure}

\begin{figure}
\centering
\includegraphics[width=0.6\textwidth]{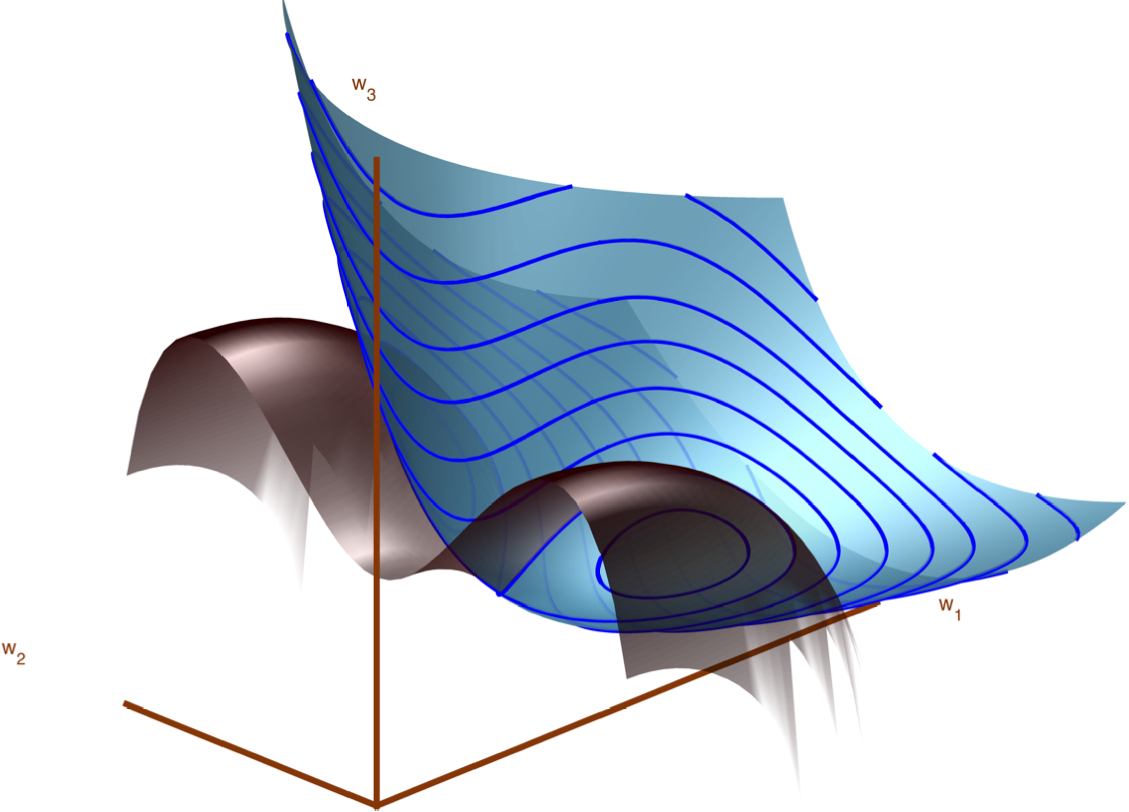}
\includegraphics[width=0.35\textwidth]{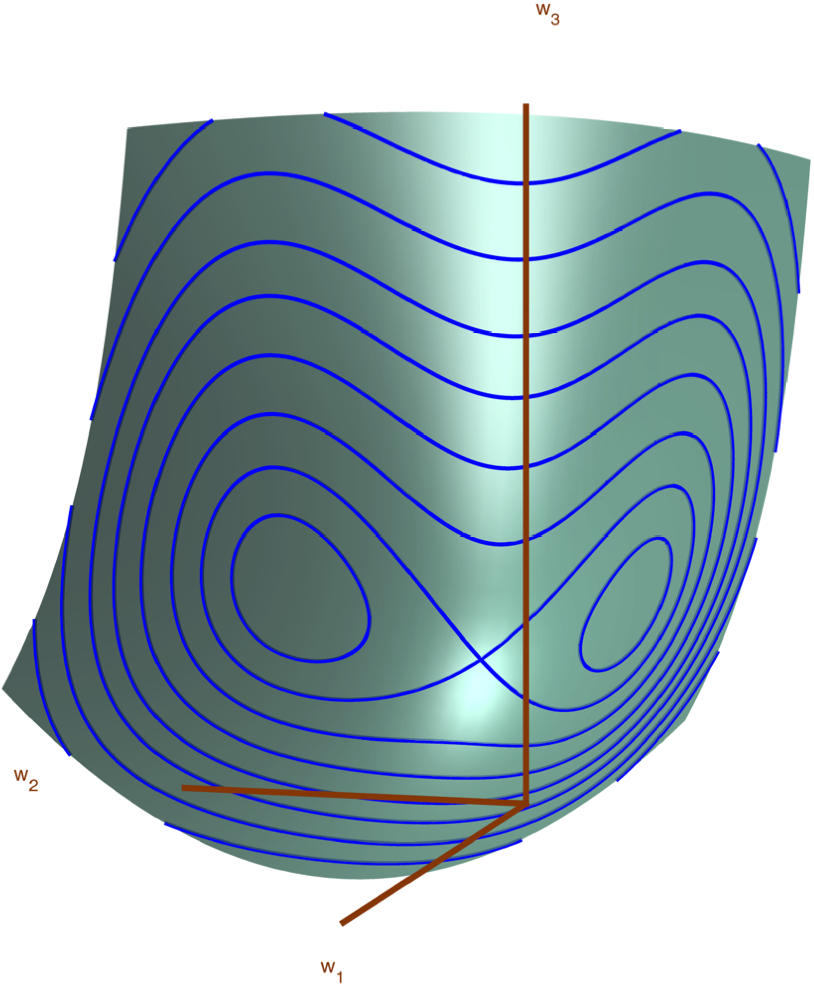}
\caption{\label{fig:homoclinic}
Two views of the phase portrait of a central force problem with homoclinic orbits.
Here $H = w_1^2+w_3^2 + w_2^4-4 w_2^2$. The energy level set $H=2$ (shown on the left)
intersects the Casimir level set $C=1$ to create two homoclinic orbits.
The situation for is similiar on other symplectic leaves: the orbit $H^{-1}(2\alpha)\cap
C^{-1}(\alpha)$ is homoclinic. The region $[0,4]\times [-2,2]\times [0,4]$ is shown.
}
\end{figure}

\begin{figure}
\centering
\includegraphics[width=0.5\textwidth]{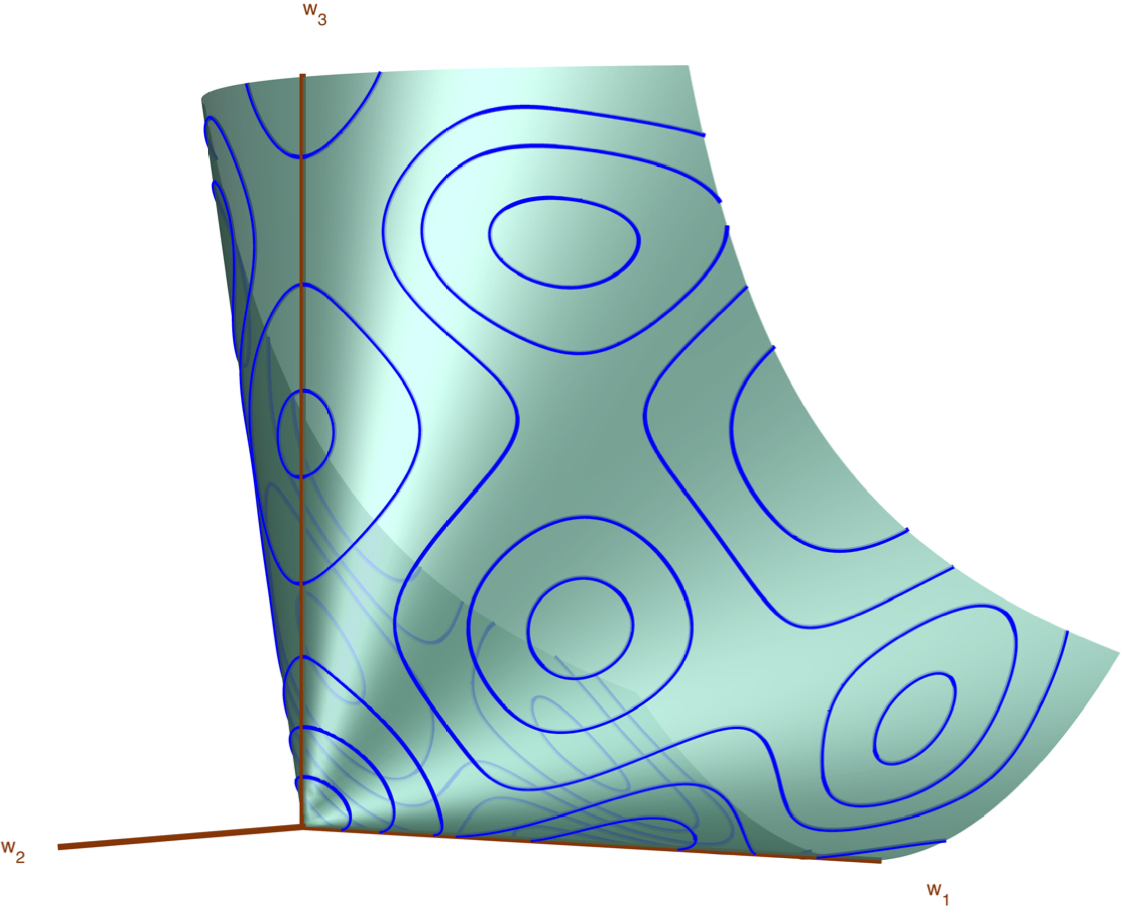}
\includegraphics[width=0.4\textwidth]{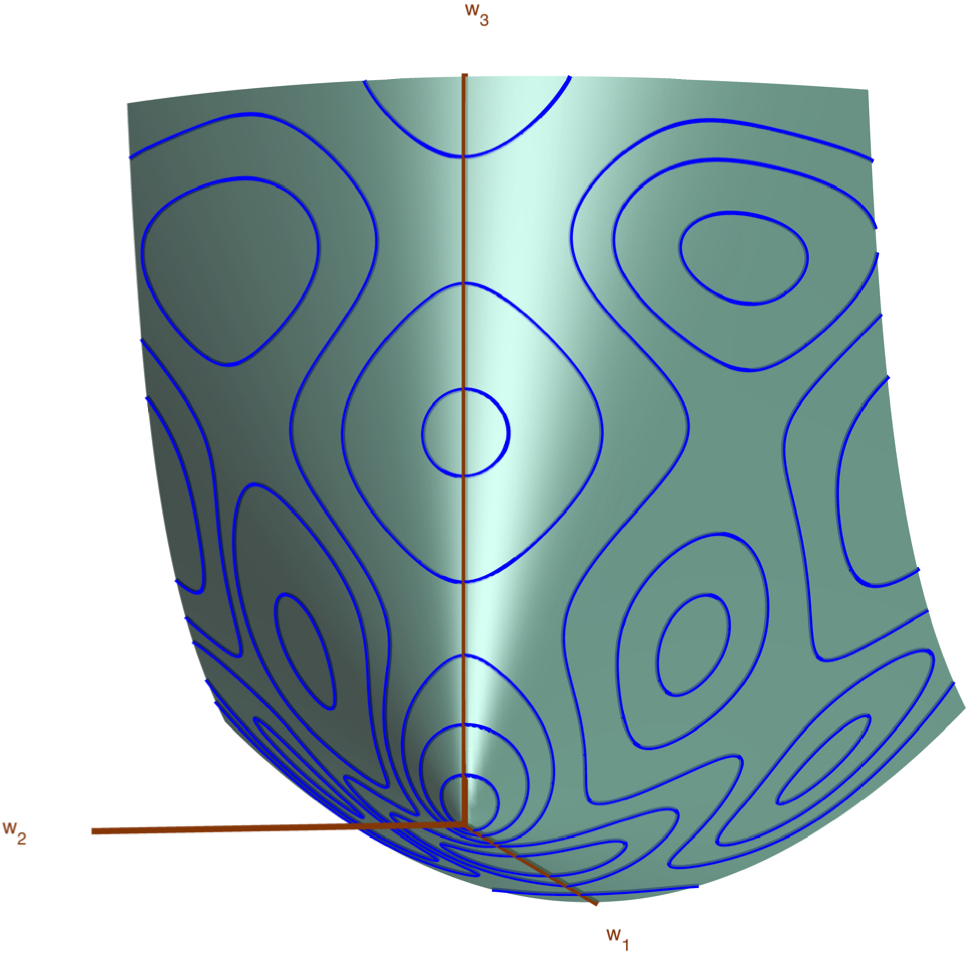}
\caption{\label{fig:complex}
Two views of the phase portrait for a more complex central force problem at zero angular momentum. 
Here $H=\sum_{i=1}^3 \cos w_i$.
The interaction between the level sets of $H$ and of the Casimir $C=w_1 w_3 - w_2^2$ creates
a complex phase portrait. The region $[0,12]\times [-6,6]\times [0,12]$ is shown.
}
\end{figure}

We have drawn such phase portraits for three systems: the Kepler problem with $\bar H=\frac{1}{2}\|p\|^2 - 1/\|q\|$
(shown in Figs.~\ref{fig:coloumb} and \ref{fig:coulomb2}); a Hamiltonian that exhibits homoclinic
orbits, $\bar H = \|q\|^2 + \|p\|^2 + (q\cdot p)^4 - 4 (q\cdot p)^2$ (shown in Fig.~\ref{fig:homoclinic}); and a Hamiltonian with 
startlingly complex phase portrait, $H=\sum_{i=1}^3 \cos w_i$ (shown in Fig.~\ref{fig:complex}).

\subsection{Reconstruction}
So far we have reduced the original set of 6 differential equations, Eq. (\ref{eq:central}), to 3, Eq. (\ref{eq:centralpoisson}),
and shown how the solutions of the reduced equations may be visualized as the level sets of the
energy and the Casimir. Strikingly, we have not yet used the conserved quantity associated 
with the symmetry. In fact, we have avoided mentioning it at all. Of course it is the angular
momentum, $J(q,p)=q\times p$. This will now make an entrance as we reconstruct the full
motion of the system.

Suppose that the reduced equations (\ref{eq:centralpoisson}) have been solved to yield a reduced
orbit $w(t)$. If this $w(t)$ is substituted into the original, unreduced, system (\ref{eq:central}), it
becomes a system of 6 linear, nonautonomous equations that reconstruct the full motion.
However, a much more dramatic simplification of the full motion is possible. This is because
the values of both $w(t)$ and of the angular momentum $J(q(t),p(t))=J(q(0),p(0))=q(0)\times p(0)$ are
known on the orbit $(q(t),p(t))$; these determine $(q(t),p(t)$ up to a single scalar function of $t$. 
This function can be found by integrating a single function of time, as follows.

Let $(\tilde q(t),\tilde p(t))$ be {\em any} curve in phase space that matches the known value
of $w(t)$---that is, such that $(\|\tilde q(t)\|^2,\tilde q(t)\cdot \tilde p(t),\|\tilde p(t)\|^2)=w(t)$. Such a curve
can be determined using only algebra. 
The desired solution, $(q(t),p(t))$, shares the same values of $w$ and $J$.
Since $w_1$, $w_2$, and $w_3$ are the defining invariants of the symmetry group, for each time $t$
there must be
an element of the symmetry group $\mathrm{O}(3)$, say $A$, that maps $(\tilde q(t),\tilde p(t))$ to $(q(t),p(t))$.
But this map also has to preserve the angular momentum $J$. 
The $\mathrm{O}(3)$ action is $(q,p)\mapsto (Aq,Ap)$,
so the action on angular momentum is $q\times p \mapsto (Aq)\times (Ap) = (\det A) A(q\times p)$. If the angular momentum is nonzero, the orthogonal matrices $A$ that preserve $q\times p$ are exactly the rotations about the axis $q\times p$. (We'll focus on the case of nonzero angular momentum; if the angular momentum is zero, then $q$ and $p$ are collinear and remain pointing in the same direction forever, and are easily determined  from $w$.)
That is, the solution can be written in the form
\begin{equation*}
\begin{aligned}
q(t) &= A(t) \tilde q(t),\\
p(t) &= A(t)\tilde p(t).
\end{aligned}
\end{equation*}
where $A(t)\in\mathrm{SO}(3)$ is some  rotation about the fixed axis $q(t)\times p(t)$.
Substituting this ansatz into the equations of motion (\ref{eq:central}) determines $A(t)$.

\setlength{\leftskip}{1.5cm}
{\medskip \small

This is an instance of a general result that the reduced motion and the conserved
momentum together determine the full motion up to an element of a certain subgroup $G_\mu$ 
of the symmetry group $G$. This is the {\em isotropy subgroup} associated with 
the value $\mu$ of the momentum map $J$. The symmetry group $G$ has a natural action on 
values of the momentum, called the coadjoint action; in this case we can write
$J(q,p)= q\times p =: \mu \in \R^3$ and the coadjoint action is $A\cdot \mu = (\det A)A\mu$. 
So in this case, the group elements that fix $\mu$ are the rotations about the axis $\mu$.

\medskip

}
\setlength{\leftskip}{0cm}

To find this  rotation $A(t)$, let $w(0)$ be the initial value of $w(t)$. We can choose coordinates on $\R^3$ so that 
\begin{equation*}
q(0) = \left(\begin{matrix}\sqrt{w_1(0)}\\ 0 \\0\end{matrix}\right),\quad
p(0) = \left(\begin{matrix}\frac{w_2(0)}{\sqrt{w_1(0)}}\\ \sqrt{w_3(0)-\frac{w_2(0)^2}{w_1(0)}} \\ 0 \end{matrix}\right)
\end{equation*}
Here the coordinates have been chosen to match the given value of $w(0)$ and so that
$\mu$ is aligned with the $z$-axis. Therefore it remains aligned with the $z$-axis for all time and the solution can
be written
\begin{equation}
\label{eq:ansatz}
q(t) = A(t)\left(\begin{matrix}\sqrt{w_1(t)}\\ 0 \\0\end{matrix}\right),\quad
p(t) = A(t)\left(\begin{matrix}\frac{w_2(t)}{\sqrt{w_1(t)}} \\ \sqrt{w_3(t)-\frac{w_2(t)^2}{w_1(t)}} \\ 0 \end{matrix}\right).
\end{equation}
where
\begin{equation*}
A(t) = \left(\begin{matrix} \cos\theta(t) & -\sin\theta(t) & 0 \\ \sin\theta(t) & \cos\theta(t) & 0 \\ 0 & 0 & 1 \end{matrix}\right)
\end{equation*}
represents the unknown rotation about the $z$-axis by an angle $\theta(t)$ which is to be determined.
(Equivalently, we can rotate the initial condition so that it lies in the $(x,y)$ plane with $q(0)$ lying on the positive
$x$-axis.)

Now inserting the ansatz (\ref{eq:ansatz}) into the equations of motion (\ref{eq:central}), and using the
reduced central force equations (\ref{eq:centralpoisson}) that are satisfied by $w(t)$, yields the reconstruction equation
\begin{equation}
\label{eq:recon}
\dot\theta = 2 \frac{\partial H}{\partial w_3} \frac{\|\mu\|}{w_1}.
\end{equation}
As the right hand side is a known function of $t$, the solution to (\ref{eq:recon}) is the phase
\begin{equation}
\label{eq:reconsol}
\theta(T) = \int_0^T 2 \frac{\partial H}{\partial w_3}(w(t))\frac{\|\mu\|}{w_1(t)}\, \mathrm{d}t.
\end{equation}
Note that the formula for the phase is not unique as it depends on the choice of coordinates in (\ref{eq:ansatz}).

Thus we see directly that the central force problem is integrable for all $H$. Orbits for which $w(t)$ is a constant
are called `relative equilibria': the full motion is periodic with period $2\pi w_1/(2 \frac{\partial H}{\partial w_3}\|\mu\|)$. Orbits
for which $w(t)$ is periodic are called `relative periodic orbits': the full motion is quasiperiodic with second period 
$2\pi/\theta(T)$ where $T$ is the period of $w(t)$. Relative periodic orbits can be seen in Figs. 
\ref{fig:coloumb}, \ref{fig:homoclinic}, and \ref{fig:complex} as closed curves in the reduced phase space.
After one period, $T$, of the reduced dynamics, the full motion does not return to its
starting point but instead picks up a phase $\theta(T)$ (in this case, a rotation in the plane
of the motion)---an example of a geometric phase,
analogous to the flip in the hammer toss experiment shown in Fig. \ref{fig:hammer}.

\subsection{Dual pairs}
\label{sec:dualpairs}
Now we come to the apparently surprising entrance of $\mathfrak{sp}(2)$, the $2\times 2$ matrices of zero trace.
The key to this is that in writing the Hamiltonian in the form $H(\|q\|^2,q\cdot p,\|p\|^2)$, we are
being given a lot of extra information about the system compared to the default situation, in which
all we know is that the Hamiltonian is invariant under a given group action. Namely, we are
being given
\begin{itemize}
\item[(i)] the reduced Hamiltonian explicitly; and 
\item[(ii)] the function
$(q,p)\mapsto (\|q\|^2,q\cdot p,\|p\|^2)$.
\end{itemize}
 In this section we will explore the consequences of this extra information.

Let $G_2$ be a matrix group with a Hamiltonian action on $T^*\R^n$ and momentum map $J_2\colon T^*\R^3\to \mathfrak{g}_2^*$. Functions of the form $H\circ J_2$ are called {\em collective Hamiltonians} and play an important
and useful role in Hamiltonian dynamics, although they seem to be much less well known than symmetric Hamiltonians and their associated tools of Noether's theorem and symplectic reduction. Collective motion is
covered in Section 28 of Guillemin and Sternberg \cite{gu-st} and Section 12.4 of Marsden \& Ratiu \cite{ma-ra}, for example. The key facts are that
\begin{itemize}
\item[(i)] $J_2$ is a Poisson map, i.e. $\{F,G\}_{\mathfrak{g}_2^*}\circ J_2 = \{F\circ J_2,G\circ J_2\}_{T^*\R^n}$ 
for all functions $F,G\colon T^*\R^n\to\R$; and 
\item[(ii)] the Hamiltonian vector field of $H\circ J_2$ on $T^*\R^n$
descends to the Hamiltonian vector field of $H$ on $\mathfrak{g}_2^*$. 
\end{itemize}
In the treatment of the 
central force problem above, we have verified these facts `by hand' for this example.

In the central force problem, $J_2=(\|q\|^2,q\cdot p,\|p\|^2)$ is the momentum map for the action of $\mathrm{Sp}(2)$
on $T^*\R^3$ given by
\begin{equation*}
\left(\begin{matrix}q \\ p \end{matrix}\right) \mapsto \left(\begin{matrix} a & b \\ c & d\end{matrix}\right)
\left(\begin{matrix}q \\ p \end{matrix}\right),\quad a d - b c = 1.
\end{equation*}
This can be checked by writing down the Hamiltonian vector fields of the components of $J_2$, which 
are the generators of this action, or by evaluating their Poisson brackets, e.g.
$\{\|q\|^2,\|p\|^2\} = 4 q \cdot p$.

Thus, the dynamics of collective Hamiltonians are exceptionally easy to reduce. The reduced system
is always  Lie--Poisson system and can be written down explicitly. In the central
force problem, not only do we have a collective Hamiltonian, but we know even more, namely that {\em all} $\mathrm{O}(3)$-invariant Hamiltonians are collective.

A second reason for the appearance of $\mathfrak{sp}(2)$ is that the matrix groups 
\begin{equation}
\begin{aligned}
&G_1 := \mathrm{O}(3),\, \quad  A\cdot(q,p) = (Aq,Ap)\\
&G_2 := \mathrm{Sp}(2), \quad  \left(\begin{matrix} a & b \\ c & d\end{matrix}\right)\cdot(q,p) = 
\left(\begin{matrix} a q + b p, c q + d p\end{matrix} \right)
\end{aligned}
\end{equation}
are {\em mutual centralizers} within the group $\mathrm{Sp}(6)$ of all symplectic linear
maps on $T^*\R^3$. That is, $G_2$ consists of those matrices in $\mathrm{Sp}(6)$
that commute with all matrices in $G_1$, and vice versa. This is an example of
a {\em dual pair}.
Amongst its remarkable consequences \cite{howe,or-ra}
 are that the momentum maps $J_1$ and $J_2$ are quadratic
functions that generate all polynomial invariants of $G_2$ and $G_1$, respectively; that $G_1$ is
a symmetry group of $J_2$ and vice versa; and that the Hamiltonian $H\circ J_2$ is
$G_1$-invariant (as we used above) and vice versa.
The dual pair was 
introduced into representation theory by Roger Howe
\cite{howe}, who gave examples of its widespread occurrence in mathematical physics, including
fundamental formalism, massless particles, classical equations (wave, Laplace, Maxwell, and Dirac) and
supersymmetry.

With this in mind, we can see another motivation for the appearance of $\mathrm{Sp}(2)$, namely
that it is a symmetry  of the angular momentum level sets, for 
\begin{equation*}
(a q + b p) \times (c q + d p) =
(a d - b c)(q\times p) = q\times p.
\end{equation*}

Remarkably, Howe \cite{howe} was able to classify the main dual pairs (the `irreducible reductive') ones). 
There are just seven families of these, with $(G_1,G_2) = (\mathrm{GL}(n,F),\mathrm{GL}(m,F))$ where the field 
$F$ is the real numbers, the complex numbers, or the quarternions;
 $(\mathrm{O}(p, q, F ), \mathrm{Sp}(2k, F ))$ (the groups preserving a Hermitian (resp. skew-Hermitian) bilinear form; this is the type that arises in the central force problem); and $(\mathrm{U} (p, q), \mathrm{U} (r, s))$.

\subsection{More bodies in more dimensions}
In our discussion of the dynamics of {\em one} body moving in {\em three} dimensions under a central force, it is easy to see that the three-dimensionality of space does not play any significant role in the reduction procedure. Similarly we can guess
that with more than one body, one should replace $(q\cdot q, q\cdot p, p\cdot p)$ by the dot products of
all pairs of $q$s and $p$s. This is in fact the case, and the resulting reduction is an application
of the Howe dual pair $(\mathrm{O}(n,\R),\mathrm{Sp}(2k,\R))$. The $\mathrm{O}(3)$-invariant motion of $k=1$ body in 
$n=3$ dimensions reduces to a Lie--Poisson system in $\mathfrak{sp}(2)$; the $\mathrm{O}(n)$-invariant motion of $k$ bodies in any dimension $n$ reduces to a Lie--Poisson system in $\mathfrak{sp}(2k)$. The resulting
structure was explored by Sadetov in 2002 \cite{sadetov}.

First, note that mechanical systems with pairwise interactions typically have Hamiltonians of the form
\begin{equation}
\label{eq:pairwise}
\bar H(q,p) = \sum_{i=1}^k \frac{1}{2 m_i} \|p_i\|^2 + \sum_{1\le i<j\le k} V(\|q_i - q_j\|^2),
\end{equation}
where the positions of the bodies are $q_1,\dots, q_k$ where $q_i\in\R^n$ and their momenta are $p_1,\dots,p_k$. 
Such systems are invariant under $\mathrm{E}(n)$, the Euclidean group of translations and rotations. 
The translation degrees of freedom, being commutative, are easy to eliminate using
center-of-mass coordinates. An explicit expression for canonical coordinates
on the reduced phase space is provided by Jacobi--Bertrand--Haretu coordinates \cite{ma-ra,meyer2,sadetov}, having been fully worked out in Spiru Haretu's 1878 PhD thesis \cite{haretu}; we'll assume this has been done so that
we are given an $\mathrm{O}(n)$-invariant Hamiltonian.

The action of $\mathrm{O}(n)$ on the positions and momenta of $k$ bodies is given by
\begin{equation*}
A \cdot(q_1,\dots,q_k,p_1,\dots,p_k) = (Aq_1,\dots,Aq_k,Ap_1,\dots,Ap_k).
\end{equation*}
The invariants of this action are the pairwise dot products of the positions and momenta, which
we collect into a map
\begin{equation*}
\varphi(q_1,\dots,q_k,p_1,\dots,p_k) = X =  \left( \begin{matrix}-M^\top & L \\ K & M\end{matrix} \right)
\end{equation*}
where the $k\times k$ matrices $M$, $L$, and $K$ are given by
\begin{equation*}
M_{ij} = q_j \cdot p_i,\quad L_{ij} = q_i\cdot q_j,\quad K_{ij} = p_i \cdot p_j.
\end{equation*}
As we can guess from the 1-body case, $\varphi$ is the momentum map 
for the action of $\mathrm{Sp}(2k)$ on $T^*\R^{n\times k}$ given by
\begin{equation*}
B\cdot(q_1,\dots,q_k,p_1,\dots,p_k) = [q_1,\dots,q_k,p_1,\dots,p_k]B.
\end{equation*}
All $\mathrm{O}(n)$-invariant Hamiltonians $\bar H$ are collective and can be written in the form $\bar H = H\circ \varphi$. 
(If the original problem was of type (\ref{eq:pairwise}), then $H$ is already known explicitly.)
Thus, their equations of motion reduce to Lie-Poisson systems 
\begin{equation*}
\label{eq:kred}
\dot X = \{X, H\}
\end{equation*}
on $\mathfrak{sp}(2k)^*$.
The Lie algebra $\mathfrak{sp}(2k)$ consists of all $2k\times 2k$ matrices of the
form $x :=\left(\begin{smallmatrix}-\alpha^\top & \beta \\ \gamma & \alpha\end{smallmatrix}\right)$
where $\alpha,\beta,\gamma$ are $k\times k$ matrices and $\beta$, $\gamma$ are symmetric.
The dual $\mathfrak{sp}(2k)^*$ can be identified with $\mathfrak{sp}(2k)$ by
the inner product $\langle X, x\rangle = \frac{1}{2}\mathrm{tr}(X^\top x)$.

The unreduced system, on $T^*\R^{n\times k}$, has dimension $2nk$. The reduced system,
(\ref{eq:kred}), has phase space a vector space of dimension $\dim\mathfrak{sp}(2k)^* = k(2k+1)$.
At first sight it is surprising that the dimension of the phase space has not been itself `reduced' when $k\ge n$. 
For example, for the 3-body problem (\ref{eq:pairwise}) we have $k=2$ and $n=2$; the original phase
space (in center-of-mass coordinates) has dimension $2nk=8$ and $\dim\mathfrak{sp}(4)^*=10$.
Some reduction!

The first point to note  is that the map $\varphi$ does not map onto all of $\mathfrak{sp}(2k)^*$. 
Its range is the positive semi-definite matrices of rank $\le n$ times 
$\left(\begin{smallmatrix}0 & I \\ -I & 0\end{smallmatrix}\right)$.
(In the 1-body example, this is the solid cone $w_1 w_3-w_2^2 \ge 0$ in $\R^3$ shown in Fig. \ref{fig:coadjoint}.) 
This submanifold has dimension $2 n k - \frac{1}{2}n(n-1)$, which {\em is} a reduction in dimension.
The `extra variables' in which the reduced system is written in (\ref{eq:kred}) is the
price we pay for embedding the reduced system, with all its complicated topology of 
symplectic leaves, in a vector space. Second, the symplectic leaves themselves have
still lower dimension. The Casimirs are $\mathrm{tr}X^{2j}$, $j=1,\dots,\lfloor n/2 \rfloor$, so 
the top-dimensional leaves have dimension $2 n k - \frac{1}{2}n(n-1) - \lfloor n/2 \rfloor$.
For $n=3$, this is $6 k - 4$, that is, symplectic reduction has lowered the dimension by 4. 

In the classic 3-body problem, the original phase space including the center of mass 
has dimension 18; removing the centre of mass drops to dimension 12; the Lie--Poisson
system (\ref{eq:kred}) has dimension 10; the image of $\varphi$ has dimension 9; and
the top-dimensional symplectic leaves have dimension 8. There are also lower-dimensional leaves
corresponding to planar motions, zero-angular-momentum motions, and motions in which
some of the bodies coincide.

\section{Discussion}
The use of invariants to work with quotients by symmetry groups is extremely well established, being
the main reason that invariants were originally introduced in the 19th century. In dynamical systems, it has  
become a powerful tool in bifurcation theory \cite{go-st-sc}, but is less well established
in geometric mechanics. Perhaps one reason for this is that while it can be powerful
in specific examples, as here, it is not universally applicable (orbits are not always
specified by the values of invariants) and the geometry of the orbits and invariants has
to be worked out in each case. Cushman and Bates \cite{cu-ba} work out many 
examples of reduction in detail in this way, for example. Olver \cite{olver} in Examples
6.43 and 6.50 treats the central force problem much as we have done here.
Lerman, Montgomery, and Sjamaar \cite{le-mo-sj} treat the case of $k$ bodies
in $n$ dimensions, noting the connection with dual pairs. The full description
of the reduction to $\mathfrak{sp}(2k)^*$ is, as far as we know, only
treated in Sadetov \cite{sadetov}. The brevity of Howe's list of dual pairs \cite{howe} 
suggests that there will not be many examples arising naturally in physics
that possess the same elegant structure as the central force problem.

\bigskip
\noindent{\bf Acknowledgements.} 
This work has received funding from the European UnionÕs Horizon 2020 research and innovation programme under the Marie Sk\l odowska-Curie grant agreement No. 691070 and from the Marsden Fund of the Royal Society of New Zealand. We would like to thank Matt Wilkins and Milo Viviani for their helpful comments.


\begin{thebibliography}{99}
\bibitem{arnold} V I  Arnold, Sur la g\'eom\'etrie diff\'erentielle des groupes de Lie de dimension infinie et ses applications \`a l'hydrodynamique des fluides parfaits, {\em  Ann. Inst. Fourier (Grenoble) \bf 16}(1) (1966) 319--361.
 
\bibitem{arnold2} V I  Arnold, {\em Mathematical methods of classical mechanics}, Springer, Berlin, 1989. 
 
\bibitem{be-or} C M Bender and S A  Orszag, {\em Advanced mathematical methods for scientists and engineers},  Macgraw Hill, 1978.
 
\bibitem{cu-ba} R H Cushman and L M Bates. Global aspects of classical integrable systems. Birkh\"auser, 2012.

\bibitem{feynman} R P Feynman, R B Leighton, and M Sands, {\em The Feynman Lecctures on Physics} vol. I, Addison--Wesley, 1965.

\bibitem{go-st-sc} M I Golubitsky, I Stewart and D Schaeffer, {\em  Singularities and Groups in Bifurcation Theory. Vol. 2}, Applied Mathematical Sciences, 69, Springer-Verlag, New York, 1988.

\bibitem{gu-st} V Guillemin and S Sternberg, {\em Symplectic Techniques in Physics},  Cambridge University Press,  Cambridge, 1990.

\bibitem{haretu} S Haretu,  Th\`eses present\`ees a la Facult\'e des Sciences de Paris, Gauthier-Villars, Paris, 1878.

\bibitem{howe} R Howe, Dual pairs in physics: harmonic oscillators, photons, electrons, and singletons, in {\em  Applications of group theory in physics and mathematical physics (Chicago 1982)}, 179--207." Lectures in Appl. Math. {\bf  21} (1985).

\bibitem{la-li} L D Landau and  Lifshitz, {\em Mechanics}, 3rd ed.,  Butterworth-Heinemann, 1976.

\bibitem{le-mo-sj} E Lerman, R Montgomery, and R Sjamaar, Examples of singular reduction,  in {\em Symplectic geometry}, London Math. Soc. Lect. Note Ser.,  {\bf 192}, Cambridge University Press, Cambridge, 1993, pp. 127--155.

\bibitem{ma-ra} J E Marsden  and T Ratiu, {\em Introduction to mechanics and symmetry: a basic exposition of classical mechanical systems}, Springer, Berlin, 2013.

\bibitem{ma-we} J Marsden and A Weinstein, Reduction of symplectic manifolds with symmetry, {\em  Reports on mathematical physics \bf 5}(1) (1974) 121--130.

\bibitem{ma-we2} J E Marsden and A Weinstein, Comments on the history, theory, and applications of symplectic reduction, in {\em Quantization of singular symplectic quotients},  Birkh\"auser, Basel, 2001, pp. 1--19.

\bibitem{meyer} K R Meyer, Symmetries and integrals in mechanics, in {\em Dynamical Systems}, M. Peixoto, ed., 259--273, Academic Press, New York., 1973.

\bibitem{meyer2} K Meyer, Periodic solutions of the $N$-body problem, {\em Lect. Notes. Math. \bf 1719}, Springer, Berlin, 1999.

\bibitem{olver} P J Olver, Applications of Lie groups to differential equations, Springer, Berlin, 2000.

\bibitem{or-ra} J-P Ortega  and T Ratiu. Momentum maps and Hamiltonian reduction, Birkh\"auser, Berlin, 2013.

\bibitem{schreiber} K U Schreiber, T Kl\"ugel, J P  Wells, R B Hurst, and A Gebauer, How to detect the chandler and the annual wobble of the earth with a large ring laser gyroscope, {\em Phys. Rev. Lett. \bf 107} (2011), 173904.

\bibitem{schwarz} G W Schwarz, Smooth functions invariant under the action of a compact Lie group, {\em Topology \bf 14} (1975) 63--68.

\bibitem{sadetov} S T Sadetov, On the regular reduction of the $n$-dimensional problem of $N+ 1$ bodies to Euler--Poincar\'e equations on the Lie algebra $sp (2N)$, {\em Regular and Chaotic Dynamics \bf 7}(3) (2002), 337--350.

\end{thebibliography}
\end{document}